\documentclass[final,11pt]{article}
\usepackage{jcappub} 
\usepackage[T1]{fontenc}
\usepackage{hyperref}
\usepackage{blindtext}
\usepackage{physics}
\usepackage{empheq}
\usepackage{amsmath}
\usepackage{amssymb}
\usepackage{enumitem}
\usepackage{float}
\newcommand{\pder}[2]{\frac{\partial #1}{\partial #2}}

\title{\boldmath Dynamics of the Mixmaster Universe in a non-commutative Generalized Uncertainty Principle framework}

\author[a,c]{Sebastiano Segreto}
\emailAdd{sebastiano.segreto@uniroma1.it}

\author[a,b]{Giovanni Montani}
\emailAdd{giovanni.montani@enea.it}

\affiliation[a]{University of Rome 'La Sapienza', P.le Aldo Moro 5, Rome, 00185,Italy}
\affiliation[b]{ENEA, Fusion and Nuclear Safety Department, C. R. Frascati, Via E. Fermi 45, Frascati (RM), 00044, Italy}
\affiliation[c]{INFN, Sezione di Roma 1, P.le Aldo Moro 2, Rome, 00185, Italy}

\abstract{
  In this work, we examine the dynamical aspects of the cosmological Mixmaster model within the framework of non-commutative generalized uncertainty principle (GUP) theories.
  The theory is formulated classically by introducing a well-defined symplectic form that differs from the ordinary one, thereby inducing a general deformation of the Poisson brackets describing a precise class of GUP theories.
  In this general setting, we first investigate the behavior of the Bianchi I and Bianchi II models using Misner variables. Then, we study the Bianchi IX model in the Mixmaster approximation, which is well-known for accurately reproducing the dynamics of the point-particle Universe approaching the cosmological singularity.
  We derive the corresponding Belinsky-Khalatnikov-Lifshitz (BKL) map and then, by selecting a specific GUP model associated with string theory, we explicitly investigate its resulting features shaped by the non-commutative GUP scheme. \\
  Our findings reveal that the chaotic and ergodic behavior typically observed in the standard BKL map, which characterizes the point-Universe's approach to the singularity, is replaced by quasi-periodic orbits in the parameter space of the theory. This corresponds to an oscillatory behavior of the Universe's scale factors, dependent on the initial conditions.}

\date{}

\begin{document}
\maketitle
\flushbottom

\section{Introduction}
\label{introduction}
  The Bianchi IX cosmology, along with the type VIII model, represents the most general space-time allowed by space homogeneity \cite{Montani:2007vu, Barrow:1981sx}. 
  The interest in this cosmology is based, on the one hand, on the behavior exhibited toward the initial singularity, characterized by chaotic dynamics \cite{BKL70, ImponenteMontani2001, CornishLevin97}, commonly known as the Mixmaster model \cite{MisnerMixUn69}, and on the other hand, on the possibility of reconciling the large-volume dynamics with that of a closed isotropic Universe \cite{KirillMon2002, Grishchuk:1975ec, Lukash}. \\
  Another reason for the model's relevance lies in its nature as a general solution to Einstein equations near the singularity. Indeed the chaotic behavior of the Bianchi IX model serves as a prototype for the evolution approaching the singularity of a generic inhomogeneous Universe \cite{Belinsky:1982pk, Kirillov93, montani1995general, BeniniMont04, Barrow:2020dap} within each small spatial region, comparable to the average causal horizon. \\
  These factors make the study of the Mixmaster model crucial for comprehending the fundamental properties of the early Universe, both from a classical perspective \cite{Montani:2007vu} and a quantum one \cite{Barca:2021qdn}.\\
  In this work we address a semiclassical treatment of the Mixmaster model within the framework of Generalized Uncertainty Principle (GUP) theories \cite{Konishi:1989wk, Amati:1988tn, Kempf:1994su, Maggiore:1993rv, Segreto:2022clx, Segreto:2024vtu}.
  Specifically, we consider $\hbar$ to be sufficiently small to maintain a classical picture while retaining a finite parameter $\beta$ deforming the Heisenberg algebra. \\
  The physical validity of this idea is based on the well-established point that the Mixmaster dynamics exists in the overlapping region of the classical and quantum evolution of the early Universe \cite{Kirillov:1997fx}. This claim is naturally reliable as the Mixmaster model is valid in the limit of arbitrarily small volume and, hence, its approach to the Planck era is implicitly assumed \cite{MisnerMixUn69, Misner:1969ae}.\\
  It is precisely in this possible co-existence of the features of the oscillatory regime and those of a quantum Universe that the motivation for adopting a GUP algebra when treating the Bianchi IX cosmology in the asymptotic limit to the initial singularity arises.
  Dealing with a classical dynamical picture, which, however, retains the morphology proper to a modified uncertainty principle, makes sense in view of a qualitative implementation of the Ehrenfest theorem \cite{Ehrenfest:1927swx}: the modified Poisson algebra provides information on the corresponding behavior of the phase space variable expectation values. This formulation - which is what it is usually meant by a semiclassical approach in this context - has the merit of giving insights into how the quantum Universe dynamics approaches a semiclassical limit and which fingerprints it leaves on the resulting dynamics.
  In other words, the present study is relevant in describing the transition phase from the quantum era to the classical evolution of the Bianchi IX cosmology.\\
  A similar approach was previously pursued in \cite{ Battisti:2008qi}, where the anisotropy degrees of freedom and their conjugate momenta were treated as degrees of freedom obeying a modified GUP algebra, while the isotropic volume-like variable played the role of time in a reduced phase space approach \cite{Arnowitt:1962hi}.\\
  In \cite{Battisti:2008qi}, through the analysis of the modified Bianchi I and Bianchi II models, some properties of the chaotic behavior of the deformed Bianchi IX model were inferred. However, the specific GUP model considered, which preserved the commutation of the generalized coordinates, i.e. the anisotropies, prevented the construction of the necessary constants of motion to induce the generalized Belinski-Khalatnikov-Lifshitz (BKL) map \cite{Misner:1969ae}.\\
  Here, we consider a class of two-dimensional non-commutative GUP theory as a framework for the semiclassical analysis of a Bianchi IX cosmology in a properly reduced phase space, where the anisotropy variables are treated as non-commutative coordinates.
  The validity and foundation of such a classical Hamiltonian formalism for these theories can be found in \cite{Bruno:2024mss}, where an extensive discussion about the symplectic structure of a certain class of GUP theories is given. \\
  We first examine the behavior of the Bianchi I cosmology in this deformed phase space, considering it as a scenario where the point-Universe is far from the potential term induced by the spatial curvature of the Bianchi IX model. We determine the conditions that generalize the Kasner circle and characterize the velocity of the point-Universe.\\
  Next, we study a Bianchi II cosmology to model the bounce of the point-Universe against the potential walls of the Mixmaster model.
  This is possible since each of them represents a Bianchi II potential wall and the Mixmaster equipotential curves in anisotropy space have equilateral triangular symmetry, making its three potential walls equivalent under a \(2\pi/3\) rotation.
  This step is crucial for defining the generalized BKL map. 
  Specifically, we identify the appropriate constants of motion for a general GUP model within the selected class, showing how the function $f$ controlling the deformation of the Poisson brackets between conjugate variables plays a significant role. This is essential for determining the reflection law of the point-Universe against the potential wall. \\
  Finally, we use these results to provide insights into the asymptotic evolution toward the initial singularity of the GUP-deformed Mixmaster model.
  To extract specific information about the dynamics of the point-Universe, we select the GUP model first presented in \cite{Kempf:1994su}, which is directly connected to string theory, reproducing the same modification of the Heisenberg Uncertainty Principle (HUP) derived from some string scattering models \cite{Konishi:1989wk} and providing a quadratic correction in the momentum variables common to a large class of GUP models.
  Since the angle between the normal to the wall and the point-Universe velocity which ensures the bounce is always greater than $\pi/3$ and approaches $\pi/2$ as energy increases, we conclude that an infinite sequence of bounces toward the singularity occurs in this specific deformation setting. \\
  The primary contribution of this study is the precise characterization of the BKL map iteration. Direct numerical analysis reveals interesting and unexpected results: although the bouncing of the point-Universe against the potential walls is endless, it does not exhibit ergodic or chaotic properties in the selected parameter space. Instead, for a given initial condition, the motion stabilizes on quasi-periodic orbits, with nearby trajectories not exponentially diverging. This resembles the quasi-periodic motion identified by Misner in \cite{Misner:1969ae}, but in our GUP scenario, it represents a general feature of the Bianchi IX asymptotic dynamics rather than an isolated case. \\
  The emerging picture is one of perpetual oscillations of the Bianchi IX scale factors toward the initial singularity, with no evidence of ergodicity or chaotic dynamics. According to the BKL conjecture, which establishes that Bianchi IX dynamics mimic the local behavior of a generic cosmological solution, our analysis suggests that, in the proposed GUP semiclassical framework, the approach of the Universe to the initial singularity could differ significantly from chaotic cosmology and more closely resemble a standard oscillatory behavior of the cosmic scale factors at each space point.

\section{Classical GUP framework}
\label{cl_GUP_framework}
  Generalized uncertainty principle (GUP) theories are quantum non-relativistic theories based on a modification of the usual Heisenberg's uncertainty principle (HUP), capable of providing an effective framework to account for an altered structure of space-time. 
  Formally, this deformed quantum framework is obtained inducing a different structure of the Heisenberg algebra, i.e. by altering the canonical commutation relations between quantum conjugate operators.
  Despite these theories are naturally formulated in a quantum framework, in \cite{Bruno:2024mss} it has been demonstrated that these theories can be consistently interpreted as classical theories as well, defined by a suitable symplectic form, compatible with the Poisson brackets inherited from the quantum commutators.
  This means that the Hamiltonian formulation for the classical dynamics of these models is available. 
  The analysis of the cosmological models here presented will be addressed at this level, i.e. in this classical deformed framework.
  The physical meaning of this formulation aims to provide an effective semiclassical theory:
  we introduce deformations of the canonical symplectic structure, whose corrections to the classical dynamics can be recognized, in the limit $\hbar \to 0$, as a manifestation of the finiteness of the Planck length. \\
  The class of GUP theories we are dealing with are described by the following fundamental Poisson brackets:
  \begin{align}
    \label{Poisson}
    \nonumber
    &\{ p_i, p_j\}=0,\\
    &\{ q_i, q_j\}= L_{ij}(q,p),\\ \nonumber  
    &\{ q_i,p_j\}=\delta_{ij}f(p).
    \end{align}
  where $f(p)$ is a smooth, strictly positive function and $L:=\{L_{ij}\}$ is a skew-symmetric matrix, whose entries are smooth functions of $(q,p)$ as well.
  From \eqref{Poisson} the 2-form $\omega$ matrix can be extracted:
  \begin{equation}
   \label{Sp-form}
    \omega_{ab}=
    \begin{pmatrix}
    0 & -\frac{1}{f}\mathrm{id}_d\\
    \frac{1}{f}\mathrm{id}_d & \frac{1}{f^2}L
    \end{pmatrix},
  \end{equation}
  In order for the 2-form \eqref{Sp-form} to be symplectic, i.e. non-degenerate and closed, the functions $f$ and $L_{ij}$ must satisfy a precise set of relations. All the details regarding this can be found in \cite{Bruno:2024mss}. Hereafter, we will assume that these relations are satisfied and that the symplectic structure of the theory is ensured. \\
  In the next sections our aim is to discuss the deformed dynamics in \emph{reduced} phase space of some relevant Bianchi cosmological models dictated by the general GUP symplectic structure in \eqref{Poisson}.
  Therefore, first we review some general features of the Bianchi models, then we describe the dynamics of the deformed Bianchi I and Bianchi II spaces, to finally arrive at the representation of the Bianchi IX behavior toward the initial singularity, i.e. the Mixmaster model.

 \section{Bianchi spaces}
 Bianchi universes are homogeneous, but generally anisotropic, cosmological spaces, whose symmetry of the $3d$ spatial hyper-surfaces is described by a Lie group, which acts simply transitively on such Riemannian manifolds.
 The general space-time metric of these models, in Misner variables, can be written as\footnote{Throughout the paper, we use natural units where $c=16\pi G=1$.}:
 \begin{equation} \label{Bianchi_metric}
     ds^2=-N(t)^2 dt^2+k_0 e^{2 \alpha}(e^{2 \gamma})_{ij}\sigma^i\sigma^j.
 \end{equation}
 Here $N(t)$ is the lapse function from the $ADM$ formulation \cite{Arnowitt:1962hi}, $k_0$ is a proper constant, the $\sigma^i$'s are \emph{1-forms} describing the symmetry Lie group of the model and $\gamma_{ij}(t)=(\gamma_{+}+\sqrt{3}\gamma_{-}, \gamma_{+}-\sqrt{3}\gamma_{-},-2\gamma_{+})$ is a diagonal traceless matrix.
 Concerning the Misner variables, $\alpha(t)$ describes the isotropic expansion (or contraction) of the Universe, while the $\gamma_{\pm}(t)$ variables represent the anisotropies, i.e. the changes in shape of the Universe itself.
 In the Hamiltonian formalism the dynamics of these models is completely described by the scalar constraint, whose general form is:
  \begin{equation} \label{Bianchi_H_constraint}
     \mathcal{H}_{Bianchi}^2=N(t)e^{3 \alpha}\left(-p^2_{\alpha}+p^2_{+}+p^2_{-}+k_0^4e^{4 \alpha}\left(V(\gamma_{\pm})-1\right)\right) \approx 0,
 \end{equation}
 where $k_0$ is the constant appearing in the metric \eqref{Bianchi_metric}, the $p's$ are the conjugate momenta to the Misner variables and $V(\gamma_{\pm})$ is a potential term determined by the spatial curvature $^3R$ of the model, which is in turn completely described by the structure constants of the symmetry Lie group. To our purposes, the constant $k_0$ is irrelevant, therefore, in the following, we will consider it to be equal to one. \\
 As anticipated above, we are going to work on a \emph{reduced} phase space. This is coherent with the effective formulation represented by GUP theories, which aims to modify the relations concerning the dynamical degrees of freedom and not directly the space-time geometry. \\
 In order to proceed in this direction, we need to solve the Hamiltonian constraint \eqref{Bianchi_H_constraint} for some suitable variable and fix the time gauge by choosing a clock variable with respect to which the dynamics of the system can be described.\\
 As usual in literature, following the ADM prescription, we can solve the constraint \eqref{Bianchi_H_constraint} for the $p_{\alpha}$ variable and select the volume-like variable $\alpha$ as a time, i.e. $\dot{\alpha}=1$. This entails fixing the lapse function $N(t)=e^{-3 \alpha}/2 p_{\alpha}$.\\
 In doing so, we obtain a reduced or physical Hamiltonian in which only the physical degrees of freedom of the system, which in our scheme are the anisotropies, appear:
 \begin{equation} \label{Bianchi_reduced_H}
     -p_{\alpha}=\mathcal{H}_{ADM}=\sqrt{p^2_{+}+p^2_{-}+e^{4 \alpha}\left(V(\gamma_{\pm})-1\right)}.
 \end{equation}
 The overall potential term is now time-dependent, being $\alpha$ our selected clock. \\
 The defined four-dimensional reduced phase space is now equipped with the generic symplectic form \eqref{Sp-form}, which induces the fundamental Poisson brackets \eqref{Poisson} and defines the general Poisson bracket between two phase space functions $F$ and $G$:
 %\begin{equation} \label{general_PB}
 %    \{F,G\}=\left(\pder{F}{q_i}\pder{G}{p_j}-\pder{F}{p_i}\pder{G}{q_j}\right)\{q_i,p_j\}+\pder{F}{q_i}\pder{G}{q_j}\{q_i,q_j\}.
 %\end{equation}
% Adapted to our setting \eqref{general_PB} becomes:
 \begin{equation} \label{general_PB}
     \{F,G\}=\left(\pder{F}{\gamma_i}\pder{G}{p_j}-\pder{F}{p_i}\pder{G}{\gamma_j}\right)\delta_{ij}f(p_{+},p_{-})+\pder{F}{\gamma_i}\pder{G}{\gamma_j}L_{ij}(\gamma_{\pm},p_{\pm}).
 \end{equation}
 This structure will dictate the (deformed) dynamics of the Bianchi models described by \eqref{Bianchi_reduced_H} that we are going to explore.

 \section{Deformed Bianchi I space}
  The Bianchi I model is the simplest among the Bianchi spaces.
  This is due to the fact that its spatial curvature, hence the potential term in \eqref{Bianchi_reduced_H}, is zero.
  The homogeneity of the space is expressed as the invariance of the metric under the Abelian group of translations $(t,x^i) \to (t, x^i+a^i)$.
  Consequently the spatial hyper-surfaces of this model, i.e. the corresponding Lie group, are identified with $\mathbb{R}^3$ and the space-time metric \eqref{Bianchi_metric} for this model reads as:
  \begin{equation}
      ds^2_{I}=-N(t)^2dt^2+e^{2 \alpha}(e^{2 \gamma})_{ij}dx^i dx^j.
  \end{equation}
 In the isotropic limit, the Bianchi I model reproduces the flat FLRW Universe.
 The reduced Hamiltonian \eqref{Bianchi_reduced_H} results to be:
  \begin{equation} \label{red_H_Bianchi_I}
      \mathcal{H}_{I}=\sqrt{p_{+}^2+p_{-}^2}
  \end{equation}
  and, by means of the general Poisson brackets \eqref{general_PB}, we can obtain the Hamiltonian equations of motion (EOM) for Bianchi I:
  \begin{equation} \label{Bianchi_I_EOM}
      \begin{cases}
          \dot{p}_{\pm}=0, \\
          \dot{\gamma}_{\pm}=\frac{p_{\pm}}{\sqrt{p^2_{+}+p^2_{-}}}f(p_{\pm}).  
      \end{cases}
  \end{equation}
 The system is easily integrable and yields the following solution:
 \begin{equation} \label{def_Bianchi_I_sol}
     \gamma_{\pm}(\alpha)=\gamma^{0}_{\pm}+\frac{p_{0\pm}}{\sqrt{p^2_{0+}+p^2_{0-}}}f(p_{0\pm}) \ \alpha
 \end{equation}
 where the quantities with 0-subscript are constants fixed by the specific Cauchy problem.
 The motion is still that of a free particle, with a trajectory deformed by the $f$ factor, moving with a deformed velocity, the modulus of which can be extracted from \eqref{Bianchi_I_EOM}:
 \begin{equation} \label{BianchiI_def_vel}
     |\dot{\gamma}|:=\sqrt{\dot{\gamma}_{+}^2+\dot{\gamma}_{-}^2}=f(p_{+},p_{-}).
 \end{equation}
 This relation marks a significant departure with respect to the standard case, where the velocity of the Bianchi I particle is equal to one, independently from the initial conditions.
 Clearly the $f$ function, in a proper limit that sets to zero the deformation, has to result to be equal to one.
 As we will see explicitly, this is achieved by introducing a suitable deformation parameter $\beta$ such that, in the limit $\beta \to 0 $, all the ordinary relations are recovered.

\section{Deformed Bianchi II space}
 The Bianchi II model is the simplest model among the Bianchi spaces with a non-zero spatial curvature.
 The symmetry Lie group is the $3d$ real Heisenberg group, which, interpreted as a smooth manifold, describes the spatial hyper-surfaces of the model. \\
 The space-time metric \eqref{Bianchi_metric}  can be written by introducing a suitable 1-form basis:
 \begin{align*}
     \sigma^1&=dx^2-x^1dx^3, \\
     \sigma^2&=dx^3, \\
     \sigma^3&=dx^1,
 \end{align*}
 and the reduced Hamiltonian of the model reads as:
 \begin{equation} \label{red_H_BianchiII}
     \mathcal{H}_{II}=\sqrt{p_{+}^2+p_{-}^2+e^{4(\alpha-2 \gamma_{+})}}.
 \end{equation}
 From here we can write down the EOM for the model:
 \begin{equation} \label{EOM_BianchiII}
     \begin{cases}
     \dot{\gamma}_{+}=\frac{p_{+}}{\mathcal{H}_{II}}f(p_{\pm}) \\
     \dot{\gamma}_{-}=\frac{p_{-}}{\mathcal{H}_{II}}+\frac{4 e^{4(2\alpha-4\gamma_{+})}}{\mathcal{H}_{II}}l(\gamma_{\pm},p_{\pm})\\
     \dot{p}_{+}=\frac{4 e^{4(2\alpha-4\gamma_{+})}}{\mathcal{H}_{II}} f(p_{\pm}) \\
     \dot{p}_{-}=0
     \end{cases}
 \end{equation}
 where we have called $l$ the only independent component of the anti-symmetric matrix $L_{ij}$.\\
 Our aim is to discuss the behavior of the particle Universe in the deformed phase space, toward the initial singularity, i.e. for $\alpha \to -\infty$.
 In this limit, the potential term in \eqref{red_H_BianchiII} can be considered as a potential wall rising very steeply.
 This allows us to describe the dynamics as that of a free particle (i.e. a Bianchi I Universe) bouncing against this potential wall.
 The condition for the potential $V$ to be relevant can be written as: 
 \begin{equation}
     \mathcal{H}_{II}^{-2}e^{4(\alpha-2 \gamma_{+})}\approx 1
 \end{equation}
 or equivalently:
 \begin{equation} \label{wall_pos}
     \gamma_{wall}:=\gamma_{+}\approx \frac{1}{2}\alpha -\ln{\mathcal{H}_{II}^2}
 \end{equation}
 where $\gamma_{wall}$ is an equipotential line in the $\gamma$-plane, allowing us to locate the position in which the potential term is dominating.\\
 A straightforward differentiation with respect to $\alpha$ reveals that, as the singularity is approached, the potential wall is receding and enables us to determine its velocity relative to the particle trajectory:
 \begin{equation} \label{wall_vel}
     \dot{\gamma}_{wall}\approx \frac{1}{2}-\frac{e^{4(\alpha-2 \gamma_{+})}}{2\mathcal{H}_{II}^2}=\frac{1}{2}\frac{p^2(\dot{\gamma})}{\mathcal{H}_{II}^2}.
 \end{equation}
 When the particle Universe is far away from the region defined by \eqref{wall_pos}, it is well-described by a Bianchi I Universe moving with velocity $|\dot{\gamma}|=f(p_{\pm})$, while from \eqref{wall_vel} we deduce that $\dot{\gamma}_{wall}=1/2$, as in the ordinary case.
 By parametrizing the velocity $\dot{\gamma}$ as $\dot{\gamma}_{+}=-f(p_{\pm})\cos\theta$ and $\dot{\gamma}_{-}=f(p_{\pm})\sin\theta$ we can establish the condition for a bounce to occur with the potential wall:
 \begin{align} \label{bounce_cond}
    & |\dot{\gamma}_{+}|>|\dot{\gamma}_{wall}| \rightarrow f(p_{\pm})|\cos\theta| >\frac{1}{2}, \nonumber \\
    & |\theta|<\cos^{-1}{\left(\frac{1}{2 f(p_{\pm})}\right)},
 \end{align}
 where the angle $\theta$ is measured with respect to the wall's perpendicular and we have dropped the modulus on $f$ since we can consider it defined as strictly positive, as explained in \cite{Bruno:2024mss}.\\
 It is clear that this limiting angle depends on the specific deformation introduced and for $f=1$ we recover the ordinary case for which $\theta<\pi/3$. \\
 When this condition is satisfied the bounce against the wall will occur, otherwise no bounce will happen and the free motion, hence the deformed Kasner epoch defined by the initial conditions, will persist indefinitely toward the singularity. \\
 To describe the bounce and to be able to determine how the Universe will emerge after it, two constants of motion are required. \\
 By inspecting the EOM \eqref{Bianchi_I_EOM} and mirroring the procedure of the standard case, two constants of motion can be found for the class of deformed theories defined by \eqref{Poisson}:
 \begin{align}
     \Omega_{1}:=p_{-}, \; \Omega_{2}:= \mathcal{H}_{II}-\frac{1}{2}\int_{\pi_{+}}^{p_{+}}\frac{1}{f(p'_{+},p_{-})}dp'_{+}.
 \end{align}
 Since these two quantities have to be conserved through the bounce, they allow us to formulate a reflection law for the motion, by means of which it is possible to determine the state of the Universe (sufficiently) after the bounce, knowing its status (sufficiently) before the bounce. \\
 Using the EOM \eqref{Bianchi_I_EOM}, which describe the relation between velocities and momenta well before and well after the bounce, and characterizing the initial state as $(\dot{\gamma}_{+})_i=-f(p_{\pm}^i)\cos\theta_i$ and $(\dot{\gamma}_{-})_i=f(p_{\pm}^i)\sin\theta_i$ and the final state as $(\dot{\gamma}_{+})_f=f(p_{\pm}^f)\cos\theta_f$ and $(\dot{\gamma}_{-})_f=f(p_{\pm}^f)\sin\theta_f$, we can finally write the general deformed reflection law, which is nothing more than a deformed BKL map for a general $f$:
 \begin{equation} \label{BKL_map_general_f}
     \begin{cases}
         \mathcal{H}_i \sin\theta_i=\mathcal{H}_f \sin\theta_f, \\
         \mathcal{H}_i-\frac{1}{2}\displaystyle \int_{{\pi_{+}}}^{p_{+}^i}\frac{1}{f(p'_{+},p_{-})}dp'_{+}=\mathcal{H}_f-\frac{1}{2} \displaystyle \int_{{\pi_{+}}}^{p_{+}^f}\frac{1}{f(p'_{+},p_{-})}dp'_{+}.
     \end{cases}
 \end{equation}
 Clearly, all the details of the map are captured by the $f$ function.\\
 The ordinary BKL map, for example, is recovered as soon as $f=1$ and combining both relations into one:
 \begin{equation} \label{BKL_map_standard}
     \sin\theta_f-\sin\theta_i=\frac{1}{2}\sin(\theta_i+\theta_f).
 \end{equation}
 In general, once a GUP model is selected, the system \eqref{BKL_map_general_f} will determine the final state of the Universe after the bounce, starting from an initial deformed Kasner epoch. This final state is the new deformed Kasner epoch to which the Universe transitions.

 \section{Deformed Mixmaster Universe}
\begin{figure}
     \centering
     \includegraphics[width=1\linewidth]{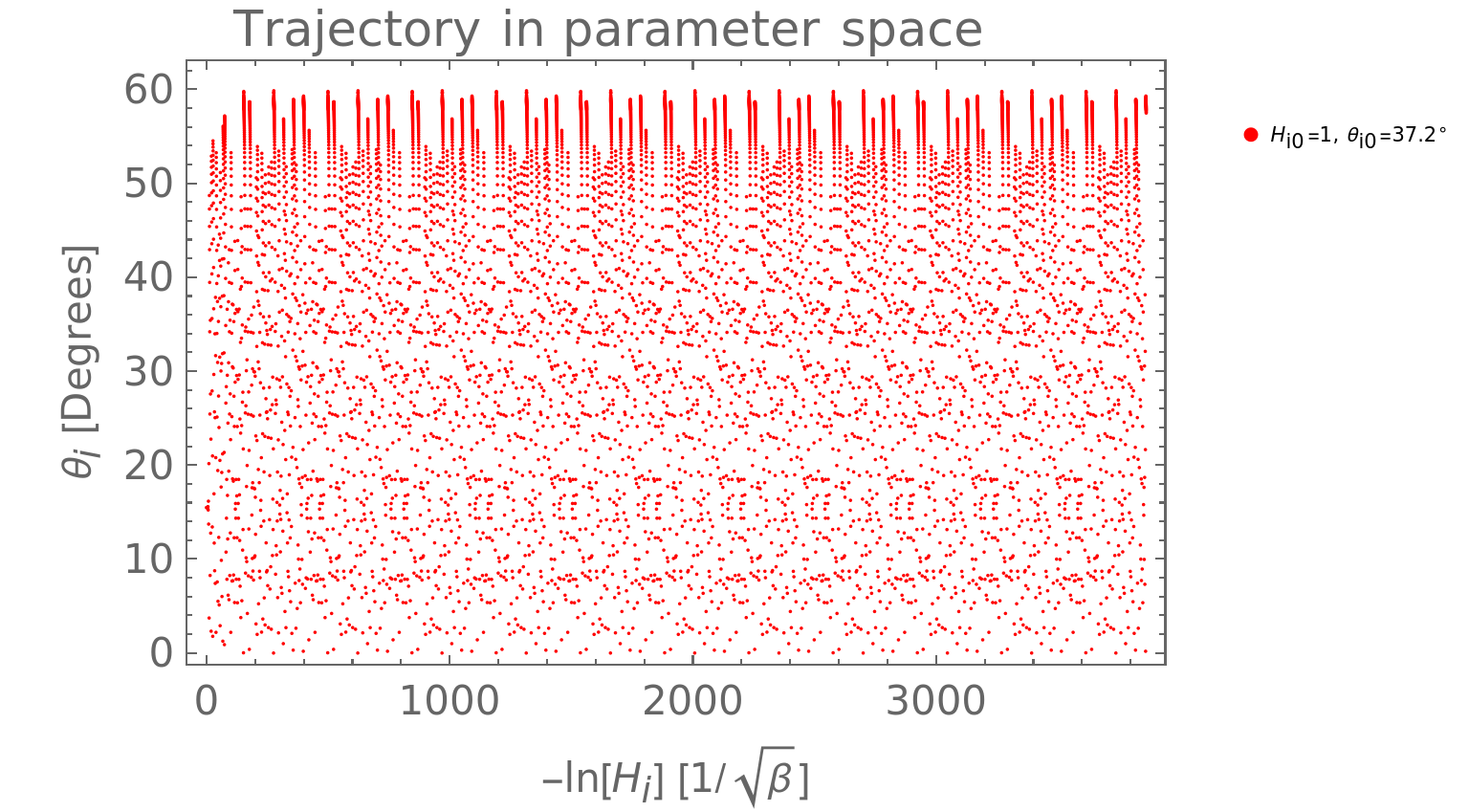}
     \caption{Trajectory of the point-Universe in the parameter space $(\mathcal{H}_i, \theta_i)$, for the initial conditions $(\mathcal{H}_i^0, \theta_i^0)$, over approximately 10000 iterations of the standard BKL map. Each point on the graph represents the state of the particle-Universe just before each collision with one of the potential walls. It is possible to observe how the system rapidly develops ergodicity, losing memory of the initial condition and chaotically exploring the entire parameter space. Note that, since in the ordinary case the energy $\mathcal{H}_i$ is decreasing at every bounce, the energy scale has been changed to a logarithmic one to improve clarity and readability.}
     \label{BKL_standard_traj}
 \end{figure}

\begin{figure}[H]
  \centering
  \includegraphics[width=1\linewidth]{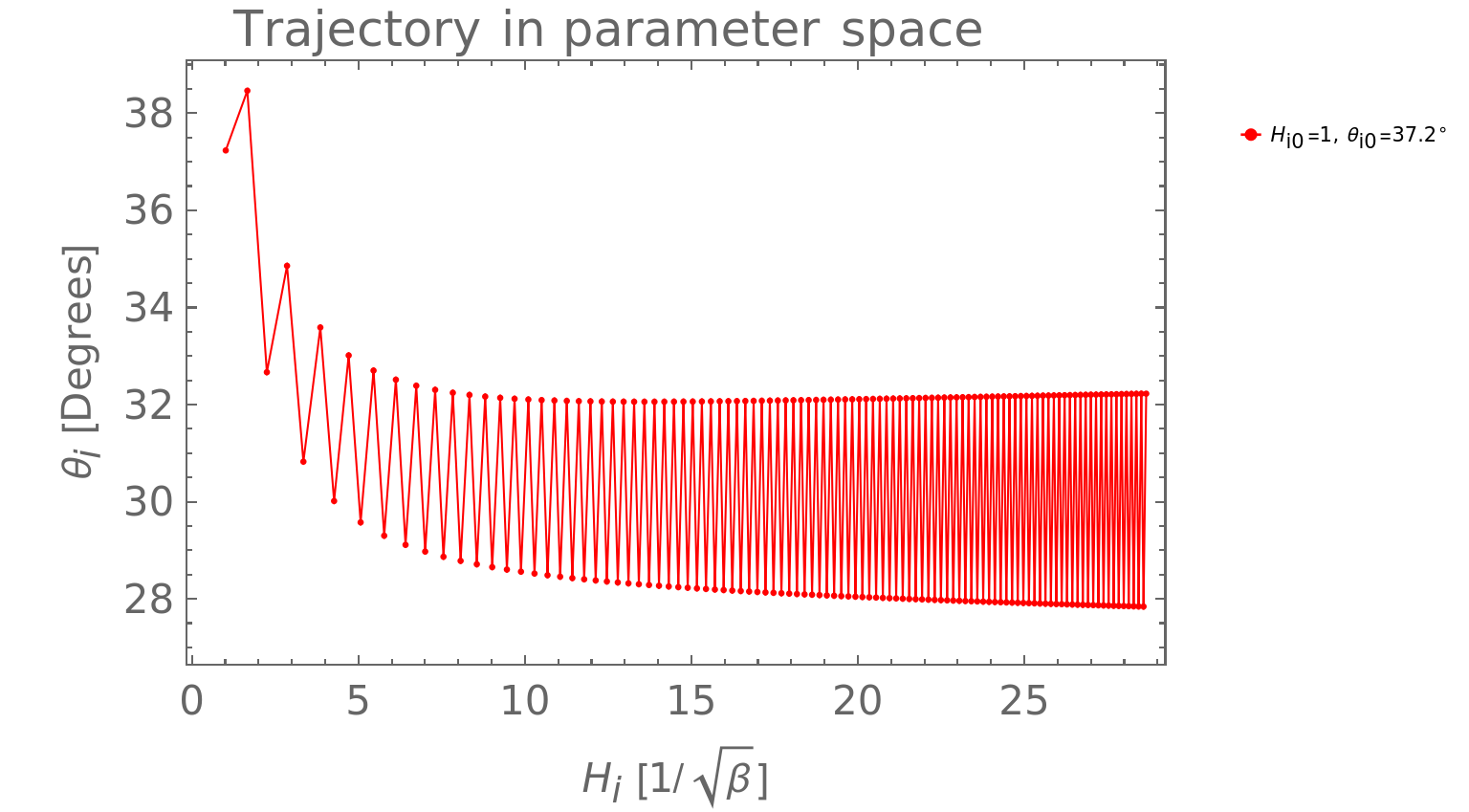}
  \caption{Orbit of the point-Universe in the parameter space ($\mathcal{H}_i$,$\theta_i$) with the initial condition ($\mathcal{H}_i^0$,$\theta_i^0$), for approximately 200 iterations of the deformed BKL map. Each point on the plot represents the state of the particle-Universe before each collision with any of the potential walls. One can observe how the system's energy increases and, more importantly, how the trajectory seems to stabilize onto a quasi-periodic orbit defined by an upper angle $\theta_{upper} \approx 32$° and a lower angle $\theta_{lower} \approx 28$°.}
  \label{gup_orb_zoom}
\end{figure}

\begin{figure}
  \centering
  \includegraphics[width=1\linewidth]{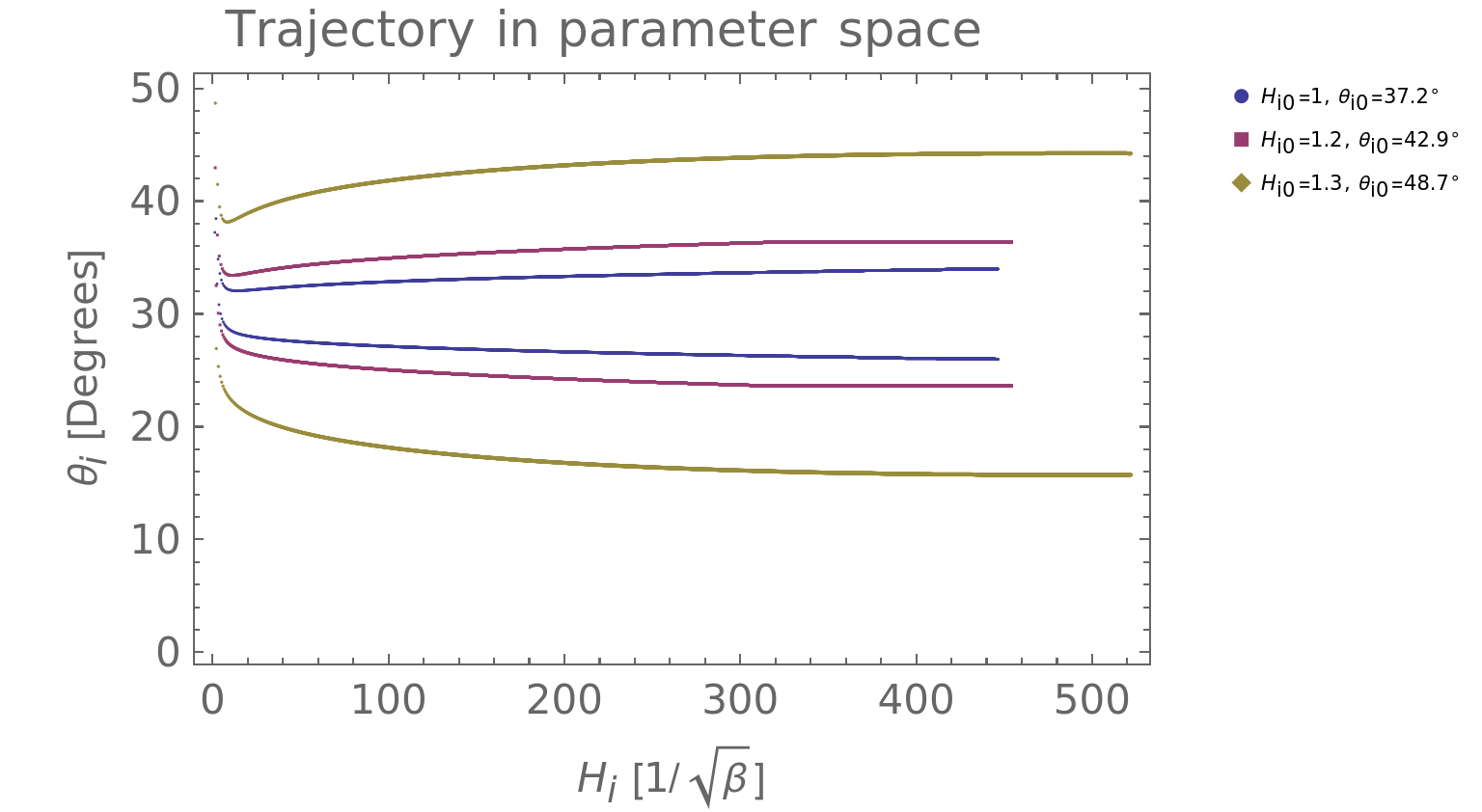}
  \caption{Plot showing various trajectories of the point-Universe in the parameter space ($\mathcal{H}_i$,$\theta_i$) for different initial conditions indicated in the legend, including those of the trajectory in Fig.\ref{gup_orb_zoom}, for approximately 500000 iterations of the deformed BKL map. The quasi-periodic tendency of the different orbits is evident, stabilizing around two boundary angles that become almost constant beyond a certain point. It is also important to note how different initial conditions affect the orbit without leading to exponentially divergent trajectories.}
  \label{gup_orb_full}
\end{figure}

 The previous analysis of Bianchi I and Bianchi II spaces enables us to develop a complete description of the deformed Bianchi IX model within the Mixmaster approximation. \\
 The symmetry Lie group of the Bianchi IX model is the $3d$ special orthogonal group $SO(3)$, with spatial hyper-surfaces identified as 3-spheres $S^3$. \\
 The space-time metric \eqref{Bianchi_metric} is described by the 1-form basis:
 \begin{align} \label{1form_B_IX}
    &\sigma^1=\sin\psi d\theta-\cos\psi \sin\theta d\phi, \nonumber \\
    &\sigma^2=\cos\psi d\theta +\sin \psi \sin\theta d\phi, \\
    &\sigma^3=-d\psi-\cos\theta d\phi, \nonumber
 \end{align}
 where $\psi,\theta,\phi$ are Euler angle coordinates on $SO(3)$, defined as usual.\\
 The 1-forms \eqref{1form_B_IX} - as every invariant 1-form for every Bianchi model - satisfy the Maurer-Cartan condition:
 \begin{equation}
     d\sigma^i=\frac{1}{2}\epsilon_{ijk}\sigma^j\wedge\sigma^k,
 \end{equation} 
 where $\epsilon_{ijk}$ are the structure  constants of the $SO(3)$ group. 
 In the isotropic limit, the Bianchi IX model corresponds to a closed FLRW Universe.\\
 The reduced Hamiltonian \eqref{Bianchi_reduced_H} is then expressed as:
  \begin{align} \label{red_H_Bianchi_IX}
    \mathcal{H}_{IX}=&\bigg\{p_{+}^2+p_{-}^2+\frac{1}{3}e^{4 \alpha}\left\{e^{-8 \gamma_{+}}-4e^{-2 \gamma_{+}}\cosh\left(2\sqrt{3}\gamma_{-}\right) \right. \nonumber \\
    & \left. +2e^{4\gamma_{+}}\left[\cosh\left(4\sqrt{3}\gamma_{-}-1\right)\right]\right\}\bigg\}^{\frac{1}{2}}.
  \end{align}
 As it is widely known, near the singularity, the Hamiltonian \eqref{red_H_Bianchi_IX} simplifies to:
    \begin{equation} \label{red_H_Mix}
    \mathcal{H}_{IX}=\sqrt{p_{+}^2+p_{-}^2+\frac{1}{3}e^{4 \alpha}\left(e^{-8 \gamma_{+}}+e^{4\gamma_{+}+4\sqrt{3}\gamma_{-}}+2e^{4\gamma_{+}-4\sqrt{3}\gamma_{-}}\right)}.
  \end{equation} 
 In this approximation, each potential term represents a Bianchi II potential, which rises very steeply in the limit $\alpha \to -\infty$. Based on the considerations made in the previous section for the Bianchi II model, this grants us the ability to treat the three potential walls in \eqref{red_H_Mix} as forming an infinite well. This well takes the shape of an equilateral curvilinear triangle with three open corners extending to infinity. Therefore, the dynamics of the point-Universe can be described by a free particle, following Bianchi I dynamics, trapped inside this receding potential well, with each bounce against the potential walls corresponding to the dynamics of the Bianchi II model. \\
 In the ordinary case, the map \eqref{BKL_map_standard} provides us with a complete description of the dynamics since, due to the symmetry of the potential, it can be iterated for every bounce.  
 Indeed, the condition for a bounce to occur is always satisfied relative to any of the three potential walls, and this will cause the point-Universe to undergo an infinite sequence of bounces. This can be easily seen from the mapping of the final angle as the new initial angle.
 Two cases can be distinguished:
 \begin{equation} \label{final_to_new_in}
     \begin{cases}
         \theta'_i=\frac{\pi}{3}-\theta_f \qquad \text{if} \quad \theta_f \leq \frac{\pi}{3}, \\
         \theta'_i=-\frac{\pi}{3}+\theta_f \qquad \text{if} \quad\theta_f > \frac{\pi}{3}.
     \end{cases}
 \end{equation}
 Therefore, since every $\theta'_i$ is always less than $\pi/3$, the ordinary condition for the bounce is always met. \\
 This implies that, as the singularity is approached, the Universe will endlessly bounce from one Kasner epoch to another, exhibiting an \emph{ergodic} and \emph{chaotic} motion.\\
 This is clearly illustrated in Fig.\ref{BKL_standard_traj}, which shows an example of a point-Universe trajectory rapidly developing ergodicity and chaos in the parameter space $(\mathcal{H}_i, \theta_i)$.\\ 
 By virtue of this behavior, this reliable approximation of the Bianchi IX model is known as Mixmaster model. \\
 Our aim in this section is to discuss the Mixmaster model in the deformed phase space defined by \eqref{Poisson}, by employing the deformed Bianchi I and Bianchi II dynamics. \\
 To conduct detailed discussions and explicit calculations, we first select a specific GUP model. \\
 We decide to examine the classical implementation of the well-known GUP theory first introduced in \cite{Kempf:1994su}, the Poisson structure 
 of which, adapted to our cosmological context, reads as:
 \begin{align}
    \label{KMM_Poisson_str}
    \nonumber
    &\{ p_i, p_j\}=0,\\
    &\{ \gamma_i, \gamma_j\}= 2\beta\left(p_{i}\gamma_{j}-p_{j}\gamma _{i}\right),\\ \nonumber  
    &\{ \gamma_i,p_j\}=\delta_{ij}\left(1+\beta \left(p_{+}^2+ p_{-}^2\right)\right).
 \end{align}
 The reasons to consider this specific model are two:
 \begin{enumerate}[label=\textit{(\roman*).}]
     \item the modification of the uncertainty principle induced by this algebra is the same obtained in some string scattering model \cite{Konishi:1989wk}; 
     \item within the class of algebras of interest in this paper, most of the possible $f$ functions, at the first order in the parameter $\beta$, reproduce the $f$ of \cite{Kempf:1994su}.
     This points out a certain universality of the quadratic correction in the momentum variables to the Heisenberg algebra. 
 \end{enumerate}
 The deformed BKL map \eqref{BKL_map_general_f} assumes the following form:
 \begin{equation} \label{BKL_map_KMM}
     \begin{cases}
         \mathcal{H}_i\sin\theta_i=\mathcal{H}_f\sin\theta_f \\
         \mathcal{H}_i+\frac{1}{2\sqrt{\beta}}\frac{\tan^{-1}\left(\frac{\sqrt{\beta}\mathcal{H}_i\cos\theta_i}{\sqrt{1+\beta \mathcal{H}_i^2\sin^2\theta_i}}\right)}{\sqrt{1+\beta \mathcal{H}_i^2\sin^2\theta_i}}=\mathcal{H}_f-\frac{1}{2\sqrt{\beta}}\frac{\tan^{-1}\left(\frac{\sqrt{\beta}\mathcal{H}_f\cos\theta_f}{\sqrt{1+\beta \mathcal{H}_f^2\sin^2\theta_f}}\right)}{\sqrt{1+\beta \mathcal{H}_f^2\sin^2\theta_f}}
     \end{cases}
 \end{equation}
 where we have used the EOM \eqref{Bianchi_I_EOM} and the deformed velocity \eqref{BianchiI_def_vel} with the explicit form of the chosen $f$.
 Note that this velocity is always greater than the ordinary one and that consistently reduces to it in the limit $\beta \to 0$. \\
 The bounce condition \eqref{bounce_cond} with respect to one of the potential walls is given by:
 \begin{equation} \label{KMM_bounce_cond}
     |\theta_i|<\cos^{-1}\left(\frac{1}{2\left(1+\beta(p_{+}^2+p_{-}^2)_i\right)}\right).
 \end{equation}
 This deformed limiting angle will depend on the initial energy for every bounce. This is due to the fact that well before the bounce itself we can set $\mathcal{H}_i^2=(p_{+}^2)_i+(p_{-}^2)_i$. 
 Most importantly, it is always greater than the ordinary one of $\pi/3$ and reduces to it in the limit $\beta \to 0$, while, at high energy, the limiting angle approaches $\pi/2$.\\
 From here and from the fact that the mapping \eqref{final_to_new_in} is still valid \footnote{A particular type of bounce, the back-bounce, is not included in this mapping. As explained in subsection \ref{backbounce} it is essentially a set of measure zero and hence can be neglected.}, we can conclude that the condition \eqref{KMM_bounce_cond} is always satisfied with respect to one of the potential walls in the deformed case as well, driving the point-Universe to be subjected to an infinite sequence of deformed bounces.\\
 All the details of this motion are explicitly contained in \eqref{BKL_map_KMM}.\\
 Inspection reveals that the final angle for each bounce depends not only on the initial angle as in \eqref{BKL_map_standard}, but on the initial energy as well.
 This means that in order to be solved, we need an initial pair $(\mathcal{H}_i,\theta_i)$ to obtain a final pair $(\mathcal{H}_f, \theta_f)$, to be used in the next step of the map.\\
 To make this more evident, the map can be expressed in a more suitable form:
 \begin{equation} \label{KMM_BKL_map_2}
     \begin{cases}
         \mathcal{H}_i+\frac{1}{2\sqrt{\beta}}\frac{\tan^{-1}\left(\frac{\sqrt{\beta}\mathcal{H}_i\cos\theta_i}{\sqrt{1+\beta \mathcal{H}_i^2\sin^2\theta_i}}\right)}{\sqrt{1+\beta \mathcal{H}_i^2\sin^2\theta_i}}=\mathcal{H}_f -\frac{1}{2\sqrt{\beta}}\frac{\tan^{-1}\left(\sqrt{\frac{\beta \mathcal{H}_f^2-\beta \mathcal{H}_i^2\sin^2\theta_i}{1+\beta \mathcal{H}_i^2\sin^2\theta_i}}\right)}{\sqrt{1+\beta \mathcal{H}_i^2\sin^2\theta_i}}\\
         \\
         \theta_f=\sin^{-1}\left(\frac{\mathcal{H}_i}{\mathcal{H}_f}\sin\theta_i\right)
     \end{cases}
 \end{equation}
 where from the first equation it is possible to obtain $\mathcal{H}_f$ and from the second $\theta_f$, given a proper initial condition.\\
 The deformed BKL map \eqref{KMM_BKL_map_2} is analytically unsolvable, necessitating numerical computations to explore its features as it approaches the singularity. Fig.\ref{gup_orb_zoom} illustrates a trajectory in the parameter space $(\mathcal{H}_i, \theta_i)$ for selected initial conditions, whereas Fig. \ref{gup_orb_full} depicts the same trajectory with many more additional iterations, along with other trajectories with different initial conditions. Each point in this space represents the state of the particle-Universe, characterized by its initial angle and its initial energy at each bounce.\\
 Analysis of these plots reveals several key observations:
 \begin{enumerate}[label=\textit{(\roman*).}]
     \item unlike the standard case, the energy of the particle-Universe increases with each bounce.
     \item After several initial bounces, trajectories in the parameter space stabilize between two distinct angles, an upper and a lower one. These angles vary slightly during the motion, in particular the upper one is slightly increasing, while the lower one is slightly decreasing. Nevertheless, they can be considered constant to all the extents after sufficient bounces.
     \item These stabilizing angles depend on the initial conditions.
     \item Slight variations in initial conditions result in slight changes in trajectories, suggesting a continuous dependence of trajectories on initial conditions.
 \end{enumerate}
 
 \noindent These findings hint at a significant conclusion regarding the dynamics of the deformed Mixmaster Universe approaching the singularity: the introduction of GUP corrections appears to remove the system's ergodicity and chaos. \\ Qualitatively, ergodicity implies the system can explore the entire configuration space regardless of initial conditions. This ability is evidently lost in the deformed dynamics, as the system remains confined within a specific region, defined by the two angles on which the motion settles, dependent on initial conditions. \\
  Chaoticity, on the other hand, qualitatively means that orbits exhibit instability and nearby trajectories diverge exponentially. This instability is absent in the deformed dynamics, as orbits demonstrate stability and nearby trajectories do not exponentially diverge in the parameter space. \\
  Although a rigorous mathematical demonstration is currently out of reach, these conclusions are strongly supported by the system's dynamics.
  The non-commutative GUP Mixmaster Universe, approaching the initial singularity, does not exhibit ergodicity or chaos but instead shows \emph{quasi-periodic orbits}, almost stabilizing between two initial angles, while the energy is increasing. 
  Undoubtedly, this marks a significant and drastic departure from the ordinary Mixmaster model.

  \subsection{Back-bounce as a set of measure zero}
  \label{backbounce}
    The deformed BKL map, being iterated, allows us to follow the motion of the point-Universe inside the potential well.
    Nevertheless, in its iteration, there is one specific type of bounce which was not taken into account and we are going to motivate why this is not necessary. \\
    This bounce potentially takes place when the particle-Universe, after a first bounce with a given wall, comes back to the previous wall from which it was coming, i.e. a back-bounce.\\
    If this happens, the new initial angle has to be mapped differently with respect to the prescription in \eqref{final_to_new_in}, specifically $\theta'_i= \pi/3+\theta_f$.
    This simple formula shows how this type of bounce is not allowed in the standard picture, but it is possible, in principle, in the deformed one, given \eqref{KMM_bounce_cond}. \\
    Nevertheless, the possibility of including in the iteration procedure of the Mixmaster map this scenario is not really feasible, since the angles $\theta_f$ for which the back-bounce is possible depend specifically on the instantaneous configuration of the bounce, i.e. they depends on the area of the triangular well and on where exactly in the wall the bounce happens.
    This means that no prescription can be given for this type of bounce within the BKL map. \\
    However, this does not impact the validity of our deformed BKL map or the features it reveals regarding the point-Universe's approach to the singularity, since this particular type of bounce can be considered negligible.
    Indeed, toward the singularity, the angle $\theta_f$ for which the back-bounce would be possible lies in a very narrow interval $[0, \epsilon]$, where $\epsilon<<1$.
    To all practical extents this means that the back-scattering angle is essentially zero.
    Being this a set of measure zero, exactly as the corners, we can neglect this possibility, without spoiling the validity of the general trend pointed out by the results of the previous section.

 \section{The method of consistent potentials as a consistency check}
  In the analysis we carried out in the previous section we have tacitly assumed that the ordinary picture, in which only one potential wall at a time is relevant, is valid in the deformed case as well.
  A possible way to verify if this assumption is valid or not is provided by the method of consistent potentials (MCP) \cite{Berger}.
  The MCP has been applied to a variety of cosmological space-times to verify if the velocity term dominated (VTD) solution is asymptotically valid in approaching the initial singularity.
  Briefly, this can be done by obtaining the VTD solution by neglecting in the Einstein equations all the spatial derivative terms and then substituting this same solution in the full Einstein equations.
  If the VTD solution is asymptotically valid and consistent, all the neglected terms will remain exponentially small. If these terms start growing, and hence they cannot be ignored, the VTD solution is not asymptotically consistent and from a certain point cease to be valid.
  This is exactly what happens in the standard Mixmaster model, where the Kasner epochs (VTD solutions) are not asymptotically consistent due to the bouncing with the potential walls.
  We of course expect that also in the deformed case the VTD solutions (deformed Kasner epochs) are not asymptotically valid. \\
  What is relevant to us in this respect is that MCP is able to point out which potential term is growing and hence is relevant for a given trajectory. \\
  We can express the Bianchi I velocity in polar coordinates $\dot{\gamma}_{+}=|\dot{\gamma}|\cos\theta$, $\dot{\gamma}_{}=|\dot{\gamma}|\sin\theta$ and rewrite the general Bianchi I solution \eqref{def_Bianchi_I_sol} as \footnote{Here the angle $\theta$ is measured counterclockwise with respect to the negative $\gamma_{+}$ axis.}:
  \begin{equation}
   \begin{aligned} \label{VTD_sol}
    &\gamma_{+}(\alpha)=|\dot{\gamma}|\cos\theta \, \alpha + \gamma_{+}^0, \\
    &\gamma_{-}(\alpha)=|\dot{\gamma}|\sin\theta \, \alpha + \gamma_{-}^0.
   \end{aligned}
  \end{equation}
  This VTD solution can now be inserted in the potential terms of \eqref{red_H_Mix} to evaluate which of them will grow with respect to a given trajectory. \\
  In the ordinary case, i.e. $f(p_{\pm})=1$ and $|\dot{\gamma}|=1$, we obtain the following characterization of the trajectories:

  \begin{align} \label{growing_region_st_MCP}
        e^{-8\gamma_{+}} \text{is growing} &\iff  -\frac{\pi}{3}<\theta<\frac{\pi}{3}  \nonumber \\
         e^{4\gamma_{+}+4\sqrt{3}\gamma_{-}} \text{is growing} &\iff  -\pi<\theta<-\frac{\pi}{3}\\
         e^{4\gamma_{+}-4\sqrt{3}\gamma_{-} } \text{is growing} &\iff  \frac{\pi}{3}<\theta<\pi  \nonumber
  \end{align}
  With the exception of the angles $\left\{-\pi/3,\pi/3, (\pm) \pi\right\}$, which identify the corners of the potential well and constitute a set of measure zero, there is no overlapping for the regions described by \eqref{growing_region_st_MCP}. This simply means that for any other angles $\theta$ only one potential wall at a time is growing. \\
  In the deformed case, i.e. $f(p_{\pm})=|\dot{\gamma}|=1+\beta(p_{+}^2+p_{-}^2)$, the MCP yields the following regions in which every potential is relevant:
  \begin{align}
    \begin{cases}
        -\cos^{-1}\left(\frac{1}{2(1+\beta p^2)}\right) < \theta < \cos^{-1}\left(\frac{1}{2(1+\beta p^2)}\right), \\
        -\pi < \theta < 2\tan^{-1}\left(\frac{\sqrt{3}(1+\beta p^2) - \sqrt{3+8\beta p^2+4\beta^2 p^4}}{p^2}\right)  \; \land \\
        2\tan^{-1}\left(\frac{\sqrt{3}(1+\beta p^2) + \sqrt{3+8\beta p^2+4\beta^2 p^4}}{p^2}\right) < \theta < \pi, \\
        -2\tan^{-1}\left(\frac{\sqrt{3}(1+\beta p^2) - \sqrt{3+8\beta p^2+4\beta^2 p^4}}{p^2}\right) < \theta < \pi \; \land \\
        -\pi < \theta < 2\tan^{-1}\left(\frac{\sqrt{3}(1+\beta p^2) + \sqrt{3+8\beta p^2+4\beta^2 p^4}}{p^2}\right).
    \end{cases}
  \end{align}
  It is not difficult to verify that these energy-dependent regions identified by the MCP overlap pairwise. \\ 
  Therefore, there will be trajectories for which two potentials at a time will be relevant.
  This means that in principle we are not entitled to use the deformed BKL map \eqref{KMM_BKL_map_2}, which has been constructed on the assumption that the dynamics can be described as a sequence of bounces against a single wall.
  A way out to recover this picture and ensuring the consistency of the result of the previous section consists in considering the ratio of the value of two of the potential walls that should be relevant for some trajectory at the same time.
  Toward the singularity, the value of any two of the potential walls of interest is actually comparable only for those orbits that fall into a very narrow region around the open corner.
  For example, considering the ratio $R$ between the second and the third potential term in \eqref{red_H_Mix}, for $\alpha<0, |\alpha|>>1$, this will be $R \propto e^{8\sqrt{3} \alpha (1+\beta p^2)\sin \theta}$.
  It is clear that under the conditions given above, $R$ will be close to the unity if and only if $\sin\theta\approx0$ in the overlapping region, which means $\theta \approx \pi$, i.e. essentially the open corner, as in the ordinary case.
  The closer we are to the singularity, the more this condition is ensured. \\
  These considerations lead us to say that even for those trajectories for which, in principle two potential walls should be considered, only one potential is really dominating to all practical extents and hence, on this ground, the picture of a particle bouncing against a single wall can be recovered in the deformed case as well, assuring the validity of the map \eqref{KMM_BKL_map_2} of the deformed theory and all the results discussed in the previous sections.

 \section{Conclusions}
  We analyzed the Mixmaster dynamics within the framework of GUP theories, where the non-commutative nature of the generalized coordinates, represented here by the anisotropy degrees of freedom, is taken into account and addressed the resulting Hamiltonian picture at a classical level \cite{Bruno:2024mss}. From a physical point of view, this is equivalent to implementing an effective approach where $\hbar \to 0$, but the deformation parameter $\beta \neq 0$.
 \\
  The development of the dynamical scheme follows the standard procedure: first, the Bianchi I model is solved to obtain the modified Kasner solution; then, the BKL map is derived by studying the Bianchi II solution and identifying its constants of motion. Finally, this comprehensive set of information is used to explore the asymptotic dynamics approaching the singularity in Bianchi IX cosmology. \\
  After deriving fundamental results for a generic GUP theory within a specific class, we selected the GUP model first introduced in \cite{Kempf:1994su}, which is notably associated with string theory.\\
  The results we obtained were quite surprising compared to the standard case, where cosmological dynamics are known to exhibit chaotic behavior \cite{BKL70, ImponenteMontani2001}. In the deformed case, the asymptotic behavior near the singularity still shows a never-ending oscillation of the cosmic scale factors of the Universe, but no evidence of chaos or ergodicity emerged. Instead, we observe quasi-periodic motion that depends on the initial conditions.\\
  We also investigated the possibility that, since the point Universe bounces against the potential when the angle between its velocity and the normal to the wall exceeds $\pi/3$, two different walls could simultaneously interact with the point-Universe. We demonstrated that such a situation does occur, but only one of the two potential terms is significant at a time, as it asymptotically dominates the other. This confirms that our results are valid for the general evolution of the Mixmaster model, except for scenarios with measure zero in the initial condition space.\\
  Since the Bianchi IX cosmology near the cosmological singularity serves as a prototype for the local behavior of generic inhomogeneous models, we infer that the present results could have broader validity. We can think that, each local region having the size of the average cosmological horizon, is characterized by the same quasi-periodic oscillations. Furthermore, as the parameters characterizing these oscillations seem to vary continuously with initial conditions, without any sign of instability, the global picture of a generic inhomogeneous Universe can be described by a smooth spatial variation of these quasi-oscillation features. \\
  In conclusion, this study highlights how a modified symplectic structure can significantly alter the field equations, substantially modifying the structure of the asymptotic regime toward the initial singularity. 
  To give physical value to this scenario is the fact that the Mixmaster dynamics lies very close to a Planckian regime of the dynamics, so that a modified Poisson algebra can be expected to bring reliable information on the real quantum features of a generalized Heisenberg algebra. 
  Nonetheless an important \textit{caveat} should be emphasized regarding the considerations above. 
  A critical point lies in the real possibility of constructing localized semiclassical states in the Mixmaster dynamics. 
  This achievement is securely addressed for a given interval of time, but the limit toward the singularity is typically characterized by a spreading of the states. 
  In this respect, two main comments are relevant:
  \begin{enumerate}[label=\textit{(\roman*).}]
      \item Misner, in his original quantum analysis of the Mixmaster dynamics \cite{Misner:1969ae}, argued that it is always possible to construct semiclassical states sufficiently close to the initial singularity, although this result has no direct correlation with state localization. We are led to infer that a similar result could hold for the GUP scenario as well. \\
      \item We demonstrated the existence of a quasi-periodic dynamics in the semiclassical GUP evolution of the Mixmaster model and this achievement suggests that the standard picture of wave function spreading \cite{Giovannetti:2022qje} is significantly altered in the proposed approach (see also the wave packets spreading of the free particle, discussed in \cite{Segreto:2024vtu}).
  \end{enumerate}
 In summary, while the relation between the GUP formalism and the possible \textit{classicalization} of the Universe is a topic requiring further investigation, we can reliably claim that the present study provides valuable information on the intermediate regime between the Planck era and the subsequent classical development of the Mixmaster Universe. \\
 Finally, to complete the discussion about this work, it is worth making some phenomenological considerations regarding the implications of GUP dynamics in the current cosmological context. \\
 It is widely known \cite{Kirillov:1997fx, Zeldovich:1983cr} that the Mixmaster dynamics is expected to overlap with the Planck era and to end before or during the inflationary epoch \cite{KirillMon2002, Cianfrani:2010gj, Montani:2009hju}. \\
 The pre-inflationary nature of the Mixmaster dynamics, regardless of the specific representation we consider for the symplectic structure describing it, makes it difficult to derive direct phenomenological implications and, consequently, a specific signature of its asymptotic behavior toward the cosmological singularity. This is due to the exponential expansion characteristic of an inflationary Universe \cite{Kolb:1990vq}  which significantly isotropizes local space regions, diluting their energy density. The Universe's pre-inflationary conditions are then restored through the so-called "reheating" process \cite{Weinberg:2008zzc}.
 Despite this substantial cancellation induced by inflation on the pre-existing physical conditions of the Universe, two phenomenological features that reflect the specific Heisenberg algebra governing the Mixmaster dynamics still persist in this context. \\
 First of all, the detailed features of the Mixmaster dynamics influence the value of the cosmological horizon before inflation begins. In this respect, the existence of an asymptotic quasi-oscillatory behavior, in contrast to the standard BKL chaos, leads us to suggest that the cosmological horizon scale is different — and likely smaller — than the one discussed in the original paper by C.W. Misner \cite{MisnerMixUn69}. This information provides insight into the spatial scale at which the pre-inflationary Universe could have reached thermal equilibrium. Clearly, the implications of such a detailed morphology of the early Universe enter the inflationary picture, depending on the specific physical model we use to describe the inflationary phase transition \cite{Weinberg:2008zzc, Kolb:1990vq}. \\
 A second important potential signature of the considered classical GUP algebra emerges when we attribute the same description transposed at the quantum level to the quantum fluctuations of the inflaton field during the de Sitter phase of the Universe.
 Standard description of these aspects can be found in \cite{Kiefer2016, Maniccia:2024bax}, while studies based on modified Heisenberg algebras can be found in \cite{Barca:2023epu}.
 In this context, it would be interesting to investigate the implications of the quantum GUP algebra corresponding to the deformed Poisson brackets considered here when analyzing the quantum inflaton spectrum. Although these quantum corrections are typically inaccessible to present-day experiments, the specific signature of a quantum version of the classical GUP model proposed here could provide valuable insights into the so-called "inverted hierarchy problem" \cite{Maniccia:2024bax}.
 This corresponding non-commutative GUP quantization paradigm was developed in \cite{Segreto:2024vtu} for an arbitrary number of dimensions, and, in the context of the inflaton spectrum, could enable a rigorous study of the quantum fluctuations of a complex field.
  
\bibliographystyle{JHEP}
\bibliography{gup_mixmaster.bib}

\end{document}